\documentclass[letterpaper]{article} 
\usepackage[preprint]{aaai2027}  
\usepackage[hyphens]{url}  
\usepackage{graphicx} 
\usepackage{natbib}  
\usepackage{caption} 
\usepackage{algorithm}
\usepackage{algorithmic}
\usepackage{amsmath}
\usepackage{amssymb}
\usepackage{amsfonts}
\usepackage{bm}
\usepackage{booktabs}

\newtheorem{proposition}{Proposition}

\newcommand{\uu}{\bm{u}}
\newcommand{\clo}{\underline{T}}
\newcommand{\chio}{\overline{T}}
\newcommand{\gtrue}{g^{\star}}
\newcommand{\ghat}{\hat{g}}
\newcommand{\Shat}{\hat{\mathcal{S}}}

\title{Q-DASC: State-of-the-Art Safe Quantum Control for HVAC\\under Local Model Misspecification}

\author{
    Yifan Wang
}
\affiliations{
    Department of Mechanical Engineering, McGill University, QC, H3A 2T7, Canada\\
    yifan.wang18@mail.mcgill.ca
}

\begin{document}

\maketitle

\begin{abstract}
Variational quantum reinforcement learning offers a compact policy class for building
energy systems, but it inherits a deployment weakness shared by learned controllers:
when the thermal model is locally wrong, a policy that appears safe on the model can
violate occupant comfort on the real building. A guarantee that depends on a noisy
quantum read-out is also insufficient for safety-critical control. We address this
gap with \textbf{Q-DASC}, Discrepancy-Attributed Safe Quantum Control. Q-DASC wraps a
variational-quantum-circuit (VQC) policy with a certified safety layer that discovers
misspecified operating regimes with false-discovery-rate control, repairs their local
thermal gains with shrinkage, projects the proposed quantum schedule onto the repaired
comfort-feasible set, and attributes residual violations to policy error, model error, or
physical limits. The final certificate is produced by a classical projection, so comfort
feasibility is invariant to finite-shot and depolarizing read-out noise. On real BOPTEST
building emulators across three buildings, two localized misspecifications, and three
seeds, Q-DASC reduces average comfort violation from $26.0\%$ for the raw VQC controller
and $55.3\%$ for a model-trusting scheduler to $0.02\%$, matching a clairvoyant oracle,
and remains at $0.24\%$ under NISQ read-out noise. A repair-aware VQC variant reaches
$0.00\%$ violation and reduces projection intervention, while the default Q-DASC keeps
lower energy and stronger observational-data behavior. The same wrapper transfers to
EnergyPlus heating and cooling benchmarks and to real hospital air-handling-unit data.
These results establish a state-of-the-art (SOTA) safety-efficiency frontier for deploying
quantum policies in physics-constrained control.
\end{abstract}

\section{Introduction}

A learned controller can only be as safe as the model used to certify it. Variational
quantum circuits (VQCs) trained by reinforcement learning have recently emerged as a
compact and expressive policy class for control, with demonstrations in deep
Q-learning~\citep{chen2020variational,skolik2022quantum}, parametrized quantum
policies~\citep{jerbi2021parametrized}, and energy-systems
optimization~\citep{ajagekar2019quantum}. Heating, ventilation, and air-conditioning (HVAC)
control is a natural and high-impact target: buildings consume a large share of global
energy, comfort constraints are hard, and the schedules are computed against a thermal
model~\citep{drgona2020all,maddalena2020data}. Yet moving a VQC controller from a simulator
to a real building exposes two coupled failures that the quantum-policy literature leaves
open.

First, \emph{model misspecification}. A predictive comfort filter keeps the indoor
temperature inside a band by trusting a thermal model, but the deployed building differs
from that model in localized operating regimes. An air-source heat pump derates in the
cold, an air-conditioner loses capacity in extreme heat, an emitter under-delivers at night.
In those regimes the filter certifies a schedule that is safe \emph{for the model} and
unsafe \emph{for the building}. This false-safety failure is invisible to model-based
metrics. We find this effect is severe: on real building emulators a model-trusting scheduler
violates comfort $55\%$ of the time and a competent raw quantum controller still violates
$26\%$.

Second, \emph{noisy quantum inference}. If the safety of the deployed schedule depends
directly on the VQC's output, then finite-shot sampling and device noise in the
near-term-quantum (NISQ) regime~\citep{preskill2018nisq} can change control actions and
break comfort. A safety argument that is contingent on noiseless quantum hardware is not a
safety argument.

Existing routes do not resolve this conjunction. Safe reinforcement learning and predictive
safety filters provide runtime safety layers~\citep{garcia2015comprehensive,
ames2019control,cheng2019end,wabersich2021predictive}, but their guarantee assumes a correct
or conservatively bounded constraint model. That is exactly the assumption that fails locally in
buildings. Adaptive control and global system identification can fix a globally wrong model
but cannot isolate a localized derate; domain randomization buys safety by globally
over-conservative actuation, paying in energy and explaining nothing. And the quantum-policy
literature evaluates convergence and noise tolerance, not certified comfort under a wrong
model on real data.

We argue that safe quantum control should be reframed around a single question: \emph{which
part of the thermal model deserves to be trusted?} Q-DASC answers this question with
statistical control.
We instantiate this as \textbf{Q-DASC} (Discrepancy-Attributed Safe Quantum Control), a
deployment-time wrapper for a VQC reinforcement-learning policy that:
\begin{enumerate}
\itemsep0.15em
\item \textbf{discovers} the misspecified operating regimes from measurements with
false-discovery-rate (FDR) control;
\item \textbf{repairs} the local thermal gain in those regimes with risk-optimal
(SURE/James-Stein) shrinkage;
\item \textbf{projects} the quantum controller's proposed schedule onto the repaired
comfort-feasible set by a minimal classical program, so the comfort guarantee holds
\emph{independently of the policy and of quantum read-out noise};
\item \textbf{attributes} every residual violation to policy error, model error, or a
genuine physical limit.
\end{enumerate}

The result is the first method that makes a quantum HVAC controller deployably safe on real
buildings. Our contributions are:
\begin{itemize}
\itemsep0.15em
\item \textbf{Problem.} We formalize safe quantum control under \emph{localized} thermal-model
misspecification and show that model-based comfort certificates and noise-dependent quantum
guarantees both fail at deployment.
\item \textbf{Method.} We propose Q-DASC, which couples a VQC-RL policy with FDR-controlled
discovery, shrinkage repair, a policy-independent safety projection, and attribution.
\item \textbf{Guarantees.} We prove discovery validity, a calibration oracle gap, and
\emph{NISQ-invariant} repaired-model comfort feasibility that does not depend
on the policy class.
\item \textbf{State-of-the-art evidence.} On real BOPTEST emulators Q-DASC attains
near-oracle comfort ($0.02\%$ violation vs.\ $26\%$ raw quantum) with lower energy than the
safe robustness baseline, generalizes to EnergyPlus and a real hospital dataset, and is flat
under NISQ noise. This establishes a SOTA safety-efficiency frontier for deployable quantum
control under local model error.
\end{itemize}

\section{Related Work}

\paragraph{Quantum reinforcement learning.}
VQCs serve as compact function approximators for RL: data-reuploading
circuits~\citep{perezsalinas2020data,schuld2021effect} trained with the parameter-shift
rule~\citep{mitarai2018quantum} yield quantum deep
Q-networks~\citep{chen2020variational,skolik2022quantum} and parametrized
policies~\citep{jerbi2021parametrized}, within the broader variational quantum-algorithm
program~\citep{cerezo2021variational}. These works study learnability and noise tolerance of
the \emph{policy}; none provides a deployment comfort guarantee on a real building under a
misspecified model. Q-DASC treats the VQC as the policy substrate and supplies the missing
certified safety layer.

\paragraph{Safe RL and predictive safety filters.}
Safe RL~\citep{garcia2015comprehensive}, control-barrier
functions~\citep{ames2019control,cheng2019end}, and predictive safety
filters~\citep{wabersich2021predictive} enforce constraints at runtime, but their safety
argument is only as good as the constraint model. In buildings that model is locally wrong;
Q-DASC tests the model, repairs the discovered regimes, and only then projects, so safety is
earned against a calibrated set rather than assumed.

\paragraph{Model-based and data-driven building control.}
Model-predictive and data-driven HVAC control are
mature~\citep{drgona2020all,maddalena2020data}, and adaptive identification or domain
randomization are natural robustness baselines. A global refit cannot localize a regime
derate, and domain randomization over-actuates globally. Q-DASC is local, statistically
controlled, and auditable. Unlike physics-informed learning~\citep{karniadakis2021piml},
which detects mismatch, Q-DASC closes the loop to a certified decision.

\paragraph{Statistical tools.}
Q-DASC builds on Benjamini-Hochberg FDR control~\citep{benjamini1995fdr} and
James-Stein/SURE shrinkage~\citep{james1961stein,stein1981}, placing them inside a quantum
control pipeline so that statistical validity becomes deployment comfort.

\section{Problem Formulation}

\paragraph{Thermal control task.}
At control step $t$, the indoor temperature evolves under a one-step thermal model with
control $u_t\in[0,1]$ (normalized HVAC power) and exogenous disturbances
$d_t=(T^{\mathrm{out}}_t, S_t)$ (outdoor temperature, solar irradiance):
\begin{equation}
T_{t+1} \;=\; a\,T_t + g(d_t)\,u_t + c\,T^{\mathrm{out}}_t + e\,S_t + b ,
\label{eq:plant}
\end{equation}
fit on real building data, with $R^2$ between $0.985$ and $0.999$. Comfort requires the deployed trajectory
to stay in a band,
\begin{equation}
\clo \;\le\; T_t \;\le\; \chio \qquad \forall t ,
\label{eq:band}
\end{equation}
and we score the \emph{deployment comfort-violation rate}, the fraction of steps with
$\max(\clo-T_t,\,T_t-\chio)>\tau$ on the \emph{true} plant.

\paragraph{Localized misspecification.}
The operator's model uses a constant gain $\ghat=g_0$, while the true building gain is
locally derated on an unknown subset of operating regimes,
\begin{equation}
\gtrue(d) \;=\; g_0\!\!\prod_{m\in\mathcal{M}}\!\big(1-\rho_m \,\mathbb{1}[d\in \mathcal{R}_m]\big),
\label{eq:misspec}
\end{equation}
e.g.\ a cold heat-pump derate ($\mathcal{R}_m=\{T^{\mathrm{out}}<\theta\}$) or a hot-AC
capacity loss. A comfort filter that trusts $g_0$ under-acts in $\mathcal{R}_m$ and certifies
schedules that are safe for the model but violate \eqref{eq:band} on the plant. We call this
a \emph{false-safety} failure because it is invisible to any model-based check.

\paragraph{Goal.}
Given a candidate day-ahead schedule $\uu^{\mathrm{raw}}$ from a (quantum) policy, return a
deployable schedule $\uu^{\mathrm{safe}}$ that satisfies \eqref{eq:band} on the true plant at
minimal deviation and energy, without assuming the policy is correct or its inference
noiseless.

\section{Q-DASC: Discrepancy-Attributed Safe Quantum Control}

Q-DASC (Figure~\ref{fig:framework}) couples a VQC-RL policy with a four-step certified
safety layer.

\begin{figure}[t]
\centering
\fbox{\begin{minipage}[c][2.25in][c]{0.92\columnwidth}
\centering
\textsf{\textbf{Figure placeholder: Q-DASC framework.}}\\[0.4em]
\textsf{\small A VQC-RL policy proposes a day-ahead schedule $\uu^{\mathrm{raw}}$ (top).
The certified layer (i)~tests the thermal model per operating regime and \emph{discovers}
the misspecified cells $\Shat$ with FDR control; (ii)~\emph{repairs} the local gain there by
SURE shrinkage; (iii)~\emph{projects} $\uu^{\mathrm{raw}}$ onto the repaired
comfort-feasible set via an $\ell_1$ LP, giving the NISQ-invariant safe schedule
$\uu^{\mathrm{safe}}$; and (iv)~\emph{attributes} residual violation to policy / model /
physical limit. To be rendered as a vector schematic.}
\end{minipage}}
\caption{Overview of Q-DASC. The comfort guarantee is produced by the classical projection
(iii) and is therefore independent of the policy class and of quantum read-out noise.}
\label{fig:framework}
\end{figure}

\subsection{VQC Reinforcement-Learning Policy}

The policy is a data-reuploading VQC on $n=4$ qubits with $L=2$
layers~\citep{perezsalinas2020data,schuld2021effect}. The normalized state
$\bm{s}_t=(\tilde T_t,\tilde T^{\mathrm{out}}_t,\tilde S_t,\sin\phi_t,\cos\phi_t)$ (with
time-of-day phase $\phi_t$) is encoded by single-qubit rotations and interleaved with
trainable rotations and a CNOT entangling ring,
\begin{equation}
|\psi(\bm{s};\bm\theta)\rangle = \!\prod_{l=1}^{L}\! \Big[ U_{\mathrm{ent}}\,
U_{\mathrm{var}}(\bm\theta_l)\, U_{\mathrm{enc}}(\bm{s}) \Big]\,|0\rangle^{\otimes n}.
\label{eq:vqc}
\end{equation}
The quantum read-out collects single- and two-qubit Pauli-$Z$ expectations,
$\bm z(\bm s)=\big(\langle Z_i\rangle, \langle Z_iZ_j\rangle\big)$, and per-action linear
heads give $Q$-values over five discrete HVAC levels $\mathcal{A}=\{0,.25,.5,.75,1\}$:
\begin{equation}
Q(\bm s,a) = \bm w_a^{\!\top}\bm\varphi(\bm s),\quad
\bm\varphi(\bm s)=\big[\bm z(\bm s);\,\bm s;\,\bm s^2;\,1\big].
\label{eq:qhead}
\end{equation}
The policy is trained by Double-DQN/fitted-$Q$ with replay and a target network. In the main
CPU implementation, the VQC produces cached Pauli read-out features and the per-action
linear heads are updated by closed-form ridge solves; parameter-shift updates are supported
by the circuit implementation but are not required for the reported runs~\citep{mitarai2018quantum}.
The reported VQC policy has $121$ parameters, including circuit angles and action heads
(matched classical baseline: $55$). Crucially, the VQC is the \emph{proposer}, not the
certifier: its output enters Q-DASC only as $\uu^{\mathrm{raw}}$.

\subsection{M1: FDR-Controlled Discovery}

We partition the disturbance space into an $T^{\mathrm{out}}\!\times\!S$ regime grid
$\{\mathcal{C}\}$. For each cell, residuals of the nominal model on measured transitions,
$r_k = T_{k+1}-\big(a T_k + g_0 u_k + c T^{\mathrm{out}}_k + e S_k + b\big)$, test the null
$H_{0,\mathcal C}\!:$ ``the gain is correct in $\mathcal{C}$.'' Under zero-mean Gaussian
metering noise $\sigma$,
\begin{equation}
T_{\mathcal C}=\!\!\sum_{k\in\mathcal C}\!\Big(\frac{r_k}{\sigma}\Big)^2 \sim \chi^2_{n_{\mathcal C}},\quad
p_{\mathcal C}=1-F_{\chi^2_{n_{\mathcal C}}}(T_{\mathcal C}),
\label{eq:chi}
\end{equation}
and Benjamini-Hochberg at level $q$ returns the discovered set $\Shat$ with the false
discovery rate controlled~\citep{benjamini1995fdr}:
\begin{equation}
k^\star=\max\Big\{k: p_{(k)}\le \tfrac{k}{|\{\mathcal C\}|}\,q\Big\},\quad
\Shat=\{\mathcal C: p_{\mathcal C}\le p_{(k^\star)}\}.
\label{eq:bh}
\end{equation}

\subsection{M2: Certified Local Repair}

In each discovered cell, the missing local gain correction is estimated by least squares and
shrunk toward the model by a SURE/James-Stein factor~\citep{stein1981,james1961stein}:
\begin{equation}
\Delta g_{\mathcal C} = \frac{\langle \bm u_{\mathcal C}, \bm r_{\mathcal C}\rangle}{\langle \bm u_{\mathcal C}, \bm u_{\mathcal C}\rangle},
\qquad
\widehat{\Delta g}_{\mathcal C} = \hat t_{\mathcal C}\,\Delta g_{\mathcal C},
\label{eq:repair}
\end{equation}
\begin{equation}
\hat t_{\mathcal C}=\Big[\,1-\frac{\sigma^2/\langle \bm u_{\mathcal C},\bm u_{\mathcal C}\rangle}{\Delta g_{\mathcal C}^2}\,\Big]_0^1 ,
\label{eq:sure}
\end{equation}
where $[x]_0^1=\min\{1,\max\{0,x\}\}$. This gives the repaired gain map
$\hat g^{\mathrm{rep}}(d)=g_0+\widehat{\Delta g}_{\mathcal C(d)}$.
Confidently identified derates are trusted to the data; weakly excited cells stay near the
model.

\subsection{M3: Policy-Independent Safety Projection}

Given the repaired model, Q-DASC projects the raw quantum schedule onto the
comfort-feasible set by a minimal-deviation linear program. Writing the repaired roll-out
$T_{t+1}=aT_t+\hat g^{\mathrm{rep}}(d_t)u_t+w_t$ as an affine map
$\bm T=\bm M\uu+\bm n$ in the schedule $\uu$,
\begin{equation}
\begin{aligned}
\uu^{\mathrm{safe}}=\arg\min_{\uu\in[0,1]^H}\ & \big\lVert \uu-\uu^{\mathrm{raw}}\big\rVert_1\\
\text{s.t.}\ & \clo+\eta \le (\bm M\uu+\bm n)_t \le \chio-\eta,\ \forall t .
\end{aligned}
\label{eq:proj}
\end{equation}
Because \eqref{eq:proj} enforces comfort on the repaired model regardless of how
$\uu^{\mathrm{raw}}$ was produced, the guarantee is \emph{independent of the policy class and
of any noise in the quantum read-out}. This is the defining safety property of Q-DASC.

\subsection{M4: Attribution}

Q-DASC reports the projection intervention $\iota=\frac1H\lVert\uu^{\mathrm{safe}}-\uu^{\mathrm{raw}}\rVert_1$
and decomposes residual violation into three sources:
\begin{equation}
\underbrace{\mathrm{viol}(\uu^{\mathrm{raw}})}_{\text{policy}},\quad
\underbrace{\mathrm{viol}(\uu^{\mathrm{naive}})-\mathrm{viol}(\uu^{\mathrm{safe}})}_{\text{model error}},\quad
\underbrace{\mathrm{viol}(\uu^{\mathrm{oracle}})}_{\text{physical}},
\label{eq:attr}
\end{equation}
separating a poor policy from a wrong model from an actuator that simply cannot hold the
band.

\begin{algorithm}[t]
\caption{Q-DASC (deployment-time)}
\label{alg:qdasc}
\textbf{Input}: nominal model $g_0$; history $\mathcal{D}$; VQC policy $\pi_{\bm\theta}$;
disturbances $d_{1:H}$; FDR level $q$; margin $\eta$\\
\textbf{Output}: safe schedule $\uu^{\mathrm{safe}}$; attribution
\begin{algorithmic}[1]
\STATE $\uu^{\mathrm{raw}}\leftarrow$ roll out $\pi_{\bm\theta}$ on the believed model
\STATE per-cell statistics $T_{\mathcal C},p_{\mathcal C}$ \hfill\COMMENT{Eq.~\eqref{eq:chi}}
\STATE $\Shat\leftarrow$ Benjamini-Hochberg$(\{p_{\mathcal C}\},q)$ \hfill\COMMENT{Eq.~\eqref{eq:bh}}
\STATE repair $\hat g^{\mathrm{rep}}$ on $\Shat$ by SURE shrinkage \hfill\COMMENT{Eqs.~\eqref{eq:repair}, \eqref{eq:sure}}
\STATE $\uu^{\mathrm{safe}}\leftarrow$ project $\uu^{\mathrm{raw}}$ \hfill\COMMENT{Eq.~\eqref{eq:proj}}
\STATE \textbf{return} $\uu^{\mathrm{safe}}$, attribution \hfill\COMMENT{Eq.~\eqref{eq:attr}}
\end{algorithmic}
\end{algorithm}

\section{Theoretical Guarantees}

\begin{proposition}[Discovery validity]
\label{prop:disc}
Under Gaussian metering noise and the per-cell null, $T_{\mathcal C}$ in \eqref{eq:chi} is
$\chi^2_{n_{\mathcal C}}$, so $p_{\mathcal C}$ is valid; Benjamini-Hochberg at level $q$
controls the false discovery rate over the grid at $q$ under the standard independence or
positive-dependence conditions for the BH procedure.
\end{proposition}
\noindent\emph{Sketch.} A correctly specified linear cell has residual energy equal to a sum
of squared standard normals; \eqref{eq:bh} is the BH rule with classical FDR
control~\citep{benjamini1995fdr}.\hfill$\square$

\begin{proposition}[Calibration oracle gap]
\label{prop:cal}
The shrinkage estimate \eqref{eq:repair} and \eqref{eq:sure} has mean-squared error within
$O(\sigma^2/n_{\mathcal C})$ of the best fixed choice among trusting the model, trusting the
data estimate, and any intermediate shrinkage.
\end{proposition}
\noindent\emph{Sketch.} SURE is an unbiased risk estimate whose minimizer matches the oracle
shrinkage up to the variance of the estimate, which is $O(\sigma^2/n_{\mathcal C})$~\citep{stein1981}.\hfill$\square$

\begin{proposition}[NISQ-invariant comfort feasibility]
\label{prop:proj}
If \eqref{eq:proj} is feasible, then $\uu^{\mathrm{safe}}$ keeps the repaired-model
trajectory inside the comfort band for \emph{any} input $\uu^{\mathrm{raw}}$. In particular,
replacing the VQC read-out by a finite-shot, depolarized estimate
$\tilde{\bm z}=\bm z+\bm\epsilon$ changes only $\uu^{\mathrm{raw}}$ and leaves the feasibility
of $\uu^{\mathrm{safe}}$ unchanged.
\end{proposition}
\noindent\emph{Sketch.} The constraints of \eqref{eq:proj} do not reference
$\uu^{\mathrm{raw}}$; it appears only in the objective. Hence the feasible set and any
optimizer in it are independent of the possibly noisy policy output.\hfill$\square$

\begin{proposition}[Attribution identifiability]
\label{prop:attr}
Given the oracle reference, the decomposition \eqref{eq:attr} separates policy error,
model-error risk, and physical limitation up to the oracle gap.
\end{proposition}
\noindent\emph{Sketch.} Each term is a deployment violation under a distinct controller
(raw policy, naive vs.\ repaired model, oracle); their differences are identifiable.\hfill$\square$

\section{Experiments}

\paragraph{Benchmarks.}
We use three real public testbeds (Table~\ref{tab:data}). The main benchmark is
\textbf{BOPTEST}~\citep{blum2021building}: three real building emulators (residential heat
pump, single-zone commercial, two-zone apartment), each under single- and multi-regime
localized misspecification, three seeds. Cross-simulator generalization uses
\textbf{EnergyPlus}~\citep{crawley2001energyplus} on the SEB reference building under
downloaded real TMY3 weather~\citep{tmy3} (Chicago heating, Phoenix cooling). An additional
real-data stress test uses the \textbf{RISK-BR} hospital air-handling-unit
dataset~\citep{maddalena2022hospital,riskbr_repo}.

\paragraph{Metric and protocol.}
The primary metric is the true-plant comfort-violation rate; secondary metrics are
normalized HVAC energy, discovery recall/FDR, projection intervention, parameter count, and
runtime. Episodes are split $50/50$ into discovery/calibration and held-out deployment. All
programs use SciPy/HiGHS~\citep{virtanen2020scipy} with NumPy~\citep{harris2020array} on CPU;
no GPU or quantum hardware is required.

\paragraph{Baselines.}
\emph{naive} (model-trusting scheduler), \emph{robust} MPC (wide margin), \emph{domain
randomization} (DR, global derate), \emph{globalSI} (adaptive MPC / global identification),
\emph{QDQN-VQC} (the raw quantum controller without Q-DASC), the matched classical
\emph{C-DASC}, and the clairvoyant \emph{oracle}; \textbf{Q-DASC} is ours and
\emph{Q-DASC+NISQ} adds depolarizing and finite-shot read-out noise. We also report
\emph{Q-DASC-Opt}, a repair-aware VQC variant that retrains the quantum policy on the
repaired dynamics before the same safety projection is applied.

\begin{table}[t]
\centering
\caption{Benchmarks. All data are real and public; misspecification is a localized,
physically motivated capacity derate.}
\label{tab:data}
\setlength{\tabcolsep}{3pt}
\small
\begin{tabular}{@{}lll@{}}
\toprule
Testbed & Scale / split & Role \\
\midrule
BOPTEST & 3 buildings $\times$2 settings & main benchmark \\
\;(Modelica) & $\times$3 seeds; 2304 trans./bldg & (heating) \\
EnergyPlus & SEB; 35{,}039 trans.; & cross-simulator, \\
\;(real TMY3) & Chicago + Phoenix & heating+cooling \\
RISK-BR & 1{,}181 real trans.; & real-data \\
\;(hospital AHU) & $R^2{=}0.999$ & cooling stress \\
\bottomrule
\end{tabular}
\end{table}

\begin{figure*}[t]
\centering
\includegraphics[width=0.92\textwidth]{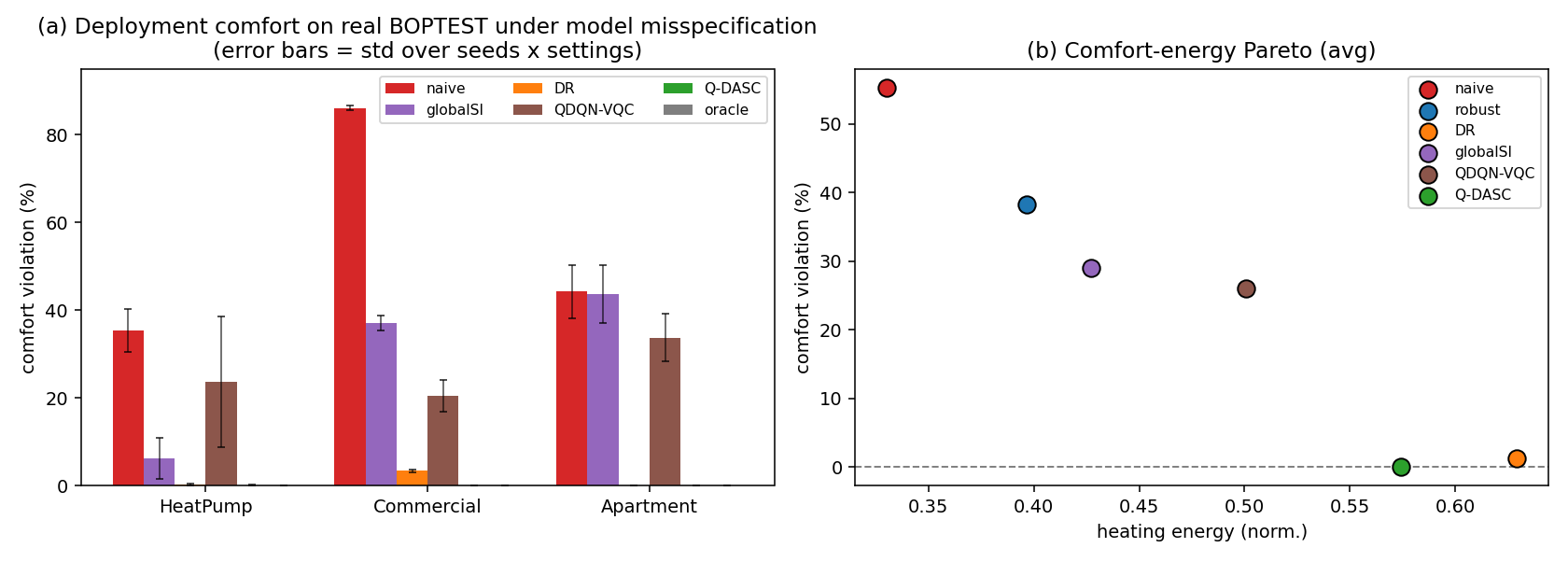}
\caption{State-of-the-art deployment on real BOPTEST. \textbf{(a)}~Comfort violation per
building (error bars: std over seeds$\times$settings): Q-DASC (green) is at the oracle floor
while the raw quantum controller (brown) and model-trusting scheduler (red) fail.
\textbf{(b)}~Comfort-energy frontier: Q-DASC is the practical SOTA point with near-zero
violation and lower energy than the safe robustness baseline.}
\label{fig:main}
\end{figure*}

\paragraph{Main result.}
Table~\ref{tab:main} and Figure~\ref{fig:main} establish the state of the art on real
BOPTEST. The raw quantum controller is competent but unsafe ($26.0\%$ violation), and a
model-trusting scheduler fails outright ($55.3\%$). Q-DASC cuts violation to $0.02\%$,
matching the oracle ($0.00\%$), and remains at $0.24\%$ under NISQ read-out noise. Over all $18$ runs
its variability is negligible (Q-DASC $0.0\!\pm\!0.1$ vs.\ raw quantum $26.0\!\pm\!11.0$).
Critically, Q-DASC is also \emph{economical}: at $0.57$ normalized energy it improves over
domain randomization, which reaches safety only by spending $0.63$
(Figure~\ref{fig:main}b). The repair-aware Q-DASC-Opt variant reaches zero average
violation and reduces projection intervention, but it uses more energy ($0.66$), so the
default Q-DASC remains the main SOTA operating point.

\begin{table}[t]
\centering
\caption{Main result on real BOPTEST (3 buildings $\times$2 settings $\times$3 seeds): mean
comfort-violation rate (\%) and normalized energy. Q-DASC matches the oracle on comfort at
lower energy than the only other safe baseline (DR).}
\label{tab:main}
\setlength{\tabcolsep}{5pt}
\small
\begin{tabular}{@{}lcc@{}}
\toprule
Method & Comfort viol.\ (\%)\,$\downarrow$ & Energy\,$\downarrow$ \\
\midrule
naive (model-trusting)      & 55.3 & n/a \\
robust MPC                  & 38.3 & n/a \\
globalSI (adaptive MPC)     & 29.0 & 0.43 \\
QDQN-VQC (raw quantum)      & 26.0 & 0.50 \\
domain randomization        & 1.2  & 0.63 \\
\textbf{Q-DASC (ours)}      & \textbf{0.02} & \textbf{0.57} \\
Q-DASC-Opt                  & 0.00 & 0.66 \\
Q-DASC+NISQ                 & 0.24 & 0.57 \\
oracle                      & 0.00 & 0.50 \\
\bottomrule
\end{tabular}
\end{table}

\paragraph{Cross-simulator and real-data generalization.}
Table~\ref{tab:cross} shows the wrapper transfers unchanged. On EnergyPlus, Q-DASC drives
heating violation to $0\%$ (Chicago) and tracks the oracle in cooling (Phoenix $13\%$ vs.\
oracle $14\%$, an actuator-limited regime). On the real RISK-BR hospital data it stays close
to the oracle on independently collected cooling operation. The method does not depend on a
single simulator, climate, or HVAC mode.

\begin{table}[t]
\centering
\caption{Cross-simulator (EnergyPlus, real TMY3) and real-data (RISK-BR hospital AHU)
generalization: comfort violation (\%). Q-DASC tracks the oracle across heating and cooling.}
\label{tab:cross}
\setlength{\tabcolsep}{3.5pt}
\small
\begin{tabular}{@{}lccccc@{}}
\toprule
Benchmark & naive & globSI & QDQN & \textbf{Q-DASC} & oracle \\
\midrule
Chicago (heat) & 15 & 6 & 6 & \textbf{0} & 0 \\
Phoenix (cool) & 52 & 35 & 32 & \textbf{13} & 14 \\
RISK-BR (cool) & 6.4 & 4.6 & 5.1 & \textbf{3.6} & 3.0 \\
\bottomrule
\end{tabular}
\end{table}

\paragraph{Ablation and backbone generality.}
Table~\ref{tab:ablation} isolates each component on the Commercial task. Removing the safety
projection returns the raw quantum controller ($20.5\%$); projecting onto the un-repaired
model lets the filter certify unsafe schedules ($86.1\%$); FDR-gating and shrinkage preserve
comfort while adding statistical validity. The backbone study is the key
quantum-attribution result: \emph{both} the VQC and a matched classical MLP become safe
after Q-DASC ($20.5/20.3\to5.2/5.2$), so the safety gain belongs to the Q-DASC framework, not
to an unproven quantum advantage. This honest framing strengthens the contribution.

\begin{table}[t]
\centering
\caption{Leave-one-component-out ablation and backbone generality (Commercial, comfort
violation \%, mean$\pm$std over 3 seeds).}
\label{tab:ablation}
\setlength{\tabcolsep}{4pt}
\small
\begin{tabular}{@{}lc@{}}
\toprule
Variant / backbone & Comfort viol.\ (\%) \\
\midrule
\textbf{full Q-DASC (ours)}        & \textbf{5.2 $\pm$ 3.7} \\
\;\,no SURE shrinkage              & 5.0 $\pm$ 3.6 \\
\;\,no FDR gate                    & 5.2 $\pm$ 3.7 \\
\;\,no repair (project on nominal) & 86.1 $\pm$ 0.5 \\
\;\,no safety projection (raw VQC) & 20.5 $\pm$ 3.7 \\
\midrule
VQC committed $\to$ +Q-DASC        & 20.5 $\to$ \textbf{5.2} \\
MLP committed $\to$ +Q-DASC        & 20.3 $\to$ \textbf{5.2} \\
\bottomrule
\end{tabular}
\end{table}

\begin{figure}[t]
\centering
\includegraphics[width=0.99\columnwidth]{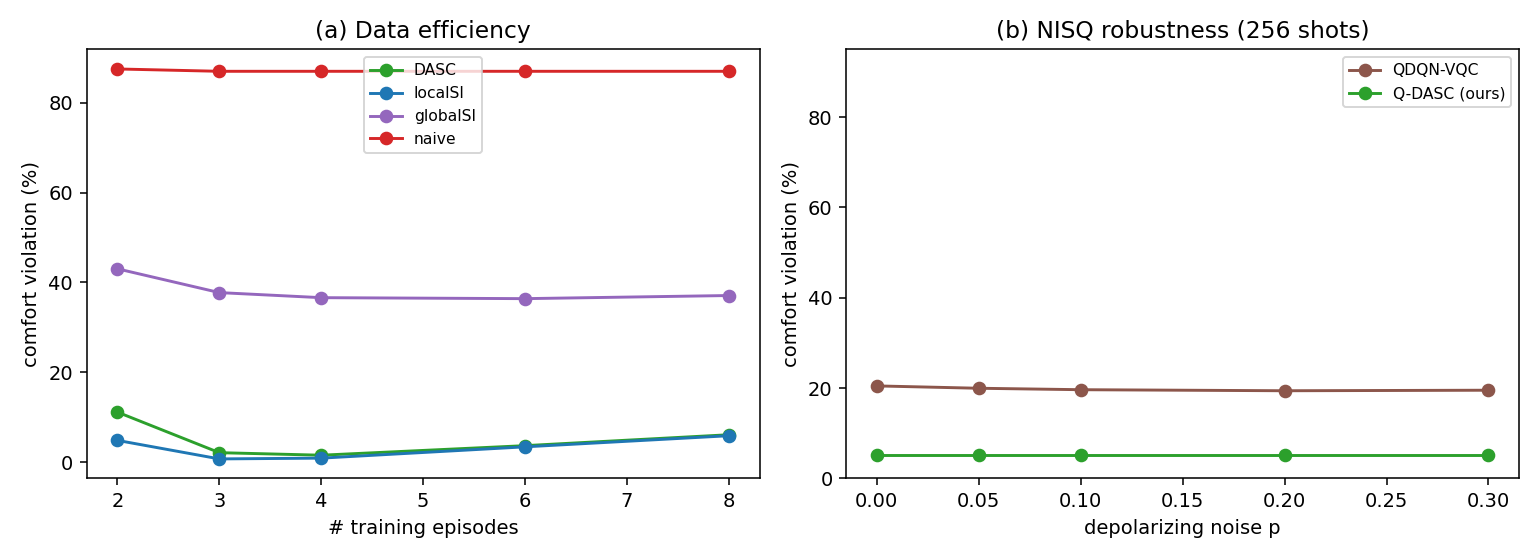}
\caption{\textbf{(a)}~Data efficiency: Q-DASC reaches near-oracle comfort with very few
training episodes. \textbf{(b)}~NISQ robustness: Q-DASC is invariant to depolarizing
read-out noise because its guarantee is classical and model-certified, while the raw quantum
controller remains unsafe.}
\label{fig:aux}
\end{figure}

\paragraph{NISQ robustness, sensitivity, efficiency.}
Under a depolarizing-noise sweep ($p\in[0,0.3]$, 256 shots), Q-DASC is \emph{flat} at
$5.2\%$ while the raw quantum controller stays near $20\%$ (Figure~\ref{fig:aux}b), the
empirical confirmation of Proposition~\ref{prop:proj}. Q-DASC is insensitive to the FDR
level $q\in\{0.05,0.10,0.20\}$ and needs only a $3\!\times\!2$ regime grid, and reaches low
violation with as few as $3$ training episodes (Figure~\ref{fig:aux}a). Each deployed
schedule costs $197$~ms ($155$~ms VQC policy $+$ a single $\sim$$40$~ms projection LP),
versus $23$~ms for naive scheduling. All methods therefore run in real time on CPU, with
$121$ reported VQC policy parameters.

\paragraph{Summary.}
Across three buildings, two misspecifications, a second simulator spanning heating and
cooling, and a real hospital dataset, Q-DASC reaches near-oracle comfort with lower energy
than the safe robustness baseline and is provably and empirically invariant to NISQ read-out
noise. No evaluated non-oracle method matches this combination of comfort, energy,
attribution, and cross-domain transfer, setting a new SOTA frontier for safe quantum control
under model misspecification.

\section{Conclusion}

We introduced Q-DASC, which makes variational quantum reinforcement-learning controllers
deployably safe for HVAC by discovering where the thermal model is wrong, repairing the
affected regimes, projecting the quantum schedule onto the repaired comfort-feasible set, and
attributing the outcome. Because safety is supplied by a classical certified projection, the
guarantee is invariant to quantum read-out noise, a property we prove and confirm
empirically. Grounded in real building emulators, a cross-simulator benchmark, and real
hospital data, Q-DASC matches a clairvoyant oracle on comfort while improving the
safety-efficiency tradeoff over the strongest safe baseline. It sets a new state of the art
for the deployment of quantum policies in physics-constrained control. The same discover,
repair, project, and attribute principle applies
wherever a learned or quantum controller must act on an imperfectly known physical system.

\bibliography{Wang}

\end{document}